\documentclass[pra]{revtex4-2}

\usepackage{graphicx}% Include figure files
\usepackage{dcolumn}% Align table columns on decimal point
\usepackage{bm}% bold math

\begin{document}
\title{Spectrally resolved two-photon interference in a modified Hong-Ou-Mandel interferometer}% Force line breaks with \\

\author{Baihong Li}
\email{li-baihong@163.com}
\author{Boxin Yuan}
%\email{1250542483@qq.com}
\author{Changhua Chen}
%\email{2540766305@qq.com}
\affiliation{%
Department of Physics, Shaanxi University of Science and Technology, Xi’an 710021, China
}%
\author{Xiao Xiang}%
%\email{xiangxiao@ntsc.ac.cn}
\author{Runai Quan}%
%\email{quanrunai@ntsc.ac.cn}
\author{Ruifang Dong}%
 \email{dongruifang@ntsc.ac.cn}
\author{Shougang Zhang}%
%\email{szhang@ntsc.ac.cn}
\affiliation{%
Key Laboratory of Time and Frequency Primary Standards, National Time Service Center, Chinese
Academy of Sciences, Xi’an 710600, China
}%

\author{Rui-Bo Jin}%
 \email{jin@wit.edu.cn}
\affiliation{%
 Hubei Key Laboratory of Optical Information and Pattern Recognition, Wuhan Institute of Technology,
Wuhan 430205, China
}%

\begin{abstract}
A modified Hong--Ou--Mandel (HOM) interference reveals that the two-photon interference phenomenon can be explained only by the concept of a two-photon wave packet rather than a single-photon one. However, the temporal interferogram in the modified HOM interferometer becomes flat in some cases so that no useful information can be extracted from time-domain measurement. Here, we theoretically explore such temporal interferogram from the frequency domain and obtain the spectrally resolved interference with high visibility. The result represents a modulation of the joint spectral intensity along both the frequency sum and the frequency difference directions. This is quite different from the cases of the spectrally resolved HOM interference and N00N state interference where the modulations happened only in one direction. Moreover, we have shown that such modulations have a potential application in the generation and characterization of high-dimensional frequency entanglement.

% and compare the obtained results with the standard HOM interference and NOON state interference. These results not only show more rich spectral information that cannot be extracted from the time domain, but also shed new light on a comprehensive understanding of the two-photon interference phenomenon in the frequency domain.

\end{abstract}

%\keywords{Suggested keywords}%Use showkeys class option if keyword
                              %display desired
\maketitle

%\tableofcontents

\section{\label{sec:1}INTRODUCTION}

Hong-Ou-Mandel (HOM) interference\cite{PRL1987}, a well-known quantum interference phenomenon, has many important applications in the fields of quantum information science such as quantum communication, quantum computation, quantum metrology \cite{RPP2021}, and remote sensing\cite{PRR2022}. A modified HOM interferometer, where an unbalanced Mach-Zender (MZ) interferometer was placed in one arm of a standard HOM interferometer, was proposed to reveal that a two-photon interference phenomenon can be explained only by the concept of a two-photon wave packet rather than a single-photon one\cite{PRA1998}. Also, it has been shown that two-photon interference only happens between two indistinguishable two-photon amplitudes instead of two photons\cite{PRL1996}.

In two-photon interference experiments, the measurements were generally performed in the time domain where the spectral information of the involved photons was integrated and lost during the measurement. With the development of high-resolved spectral technology at single-photon level, more rich phenomenons can be found in spectrally resolved two-photon\cite{SR-PRA2015,SR-OE2015,SR-PR2020,SR-OE2021,SR-NOON2021,SR-2021arxiv}, three-photon\cite{3-photons2019} or four-photon interferences \cite{4-photons2021}. This allows us to observe useful spectral information with high visibility that cannot be extracted from the time domain, especially when the visibility of time-domain interference is very low or zero \cite{SR-NOON2021}. Another advantage of this technology is that the time-domain interferogram, which would have been obtained through a standard delay-scanning with long acquisition times, can be recovered fast from the spectrally resolved interference\cite{SR-PR2020}. Moreover, spectrally resolved interference can also provide a quantum enhancement on the measurement precision, for example in chromatic dispersion measurement\cite{Light-2018} and time-delays estimation\cite{SR-2021arxiv}, as well as a higher Fisher information than non-spectrally resolved results \cite{SR-2021arxiv,SR-NOON2021}.

However, spectrally resolved two-photon interference for the modified HOM interferometer has not yet been explored, to the best of our knowledge. Specifically, the middle interferogram in the modified HOM interferometer becomes flat and its visibility is zero in some cases, e.g., when the broadband pump is used. As a result, no useful information can be extracted from such a temporal interferogram. It is therefore necessary to see whether spectral information can be found through spectrally resolved interference technology.

In this paper, we first derive a general formula of coincidence count rates for the modified HOM interferometer from the frequency domain. Based on this derivation, we theoretically explore spectrally resolved two-photon interference for the modified HOM interferometer, especially for the zero delay between signal and idler photons. It is found that the coincidence probability density (CPD) depends on both the frequency sum and the frequency difference between signal and idler photons, it is thus phase-dependent. As a result, the joint spectral intensity (JSI) can be modulated by the time delay along both the frequency sum and the frequency difference directions. These results are also compared with the cases of spectrally  resolved HOM interference and N00N state interference. Furthermore, we discuss the potential application of such modulations in the generation and characterization of high-dimensional frequency entanglement.

\section{\label{sec:2}Theory of a modified HOM interference derived from the frequency domain}

\begin{figure}[th]

\begin{picture}(100,180)
\put(0,0){\makebox(-50,180){
\scalebox{0.17}[0.17]{
\includegraphics{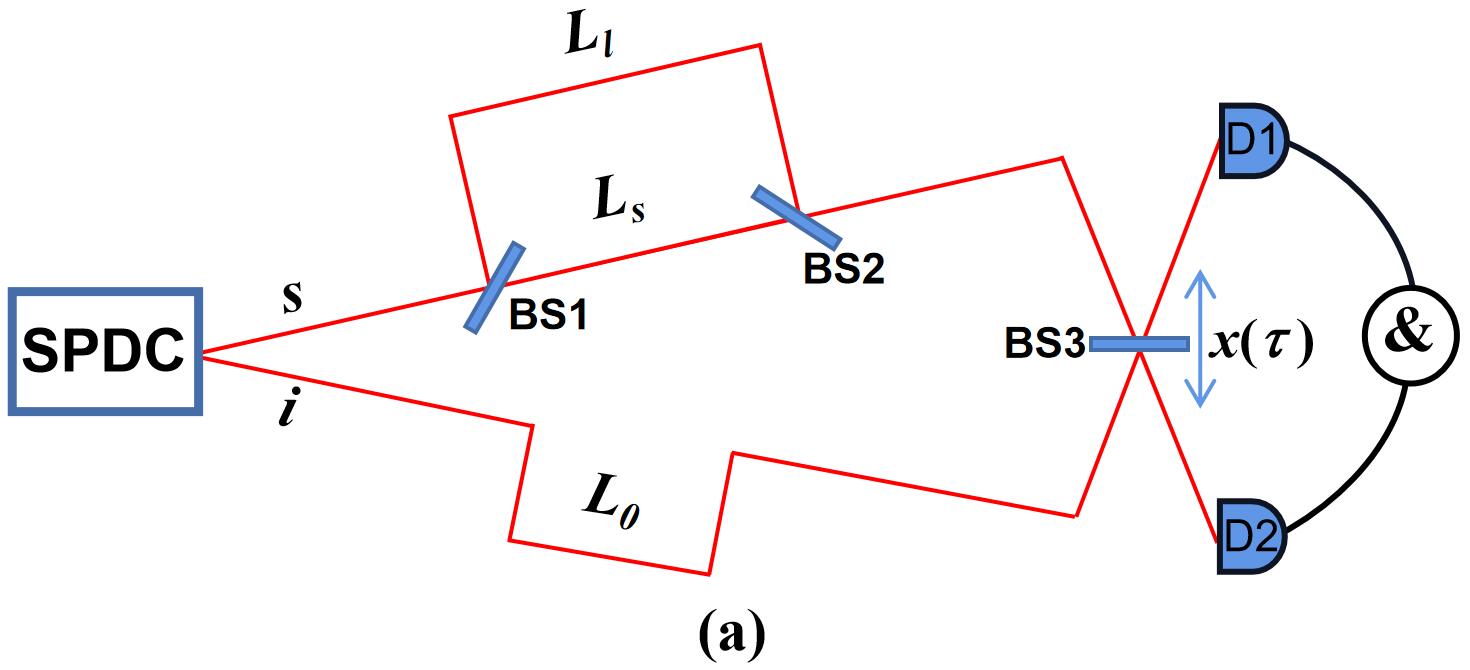}
}}}
\end{picture}\begin{picture}(100,150)
\put(0,0){\makebox(205,170){
\scalebox{0.22}[0.22]{
\includegraphics{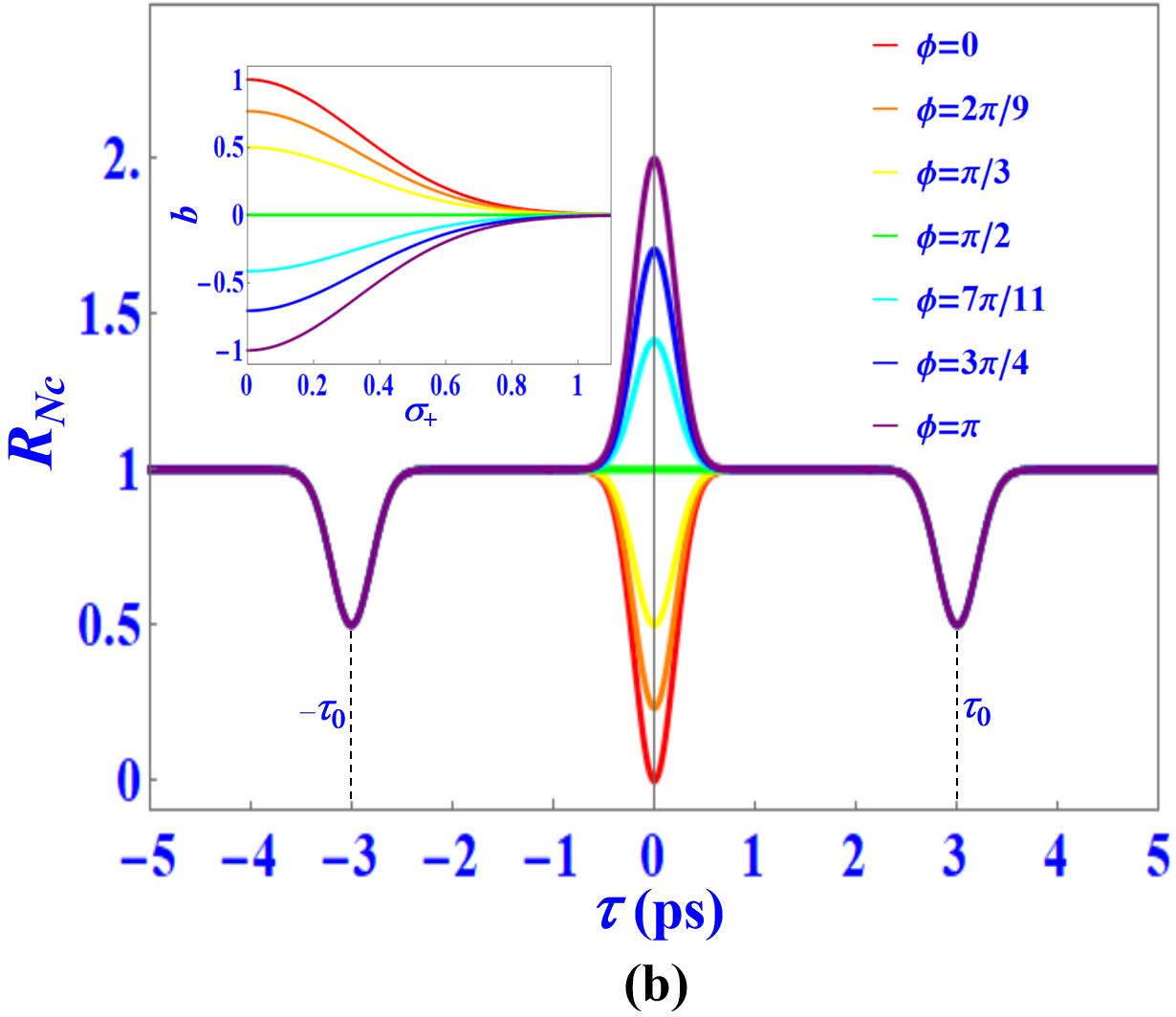}
}}}
\end{picture}
%%\vspace{-1cm}
\caption{\label{Fig1}
(a) Schematic diagram of the modified HOM interferometer. (b) The normalized interferogram for the modified HOM interferometer with different values of $\phi$. $\sigma_-=5THz$; $\tau_0=3ps$. The inset shows $b$ as functions of $\sigma_+$ and $\phi$.}
\end{figure}

To reveal more spectral information, it is necessary to derive the general expression for the modified HOM interference from the frequency domain. First, let us consider the schematic diagram of the modified HOM interferometer as shown in Fig. \ref{Fig1}(a). The entangled photons (biphotons) are generated by the spontaneous parametric down-conversion (SPDC) process. The idler ($i$) path length is $L_0$. An unbalanced MZ interferometer is placed in the signal($s$) arm, and there are two optical paths $L_l$ and $L_s$ such that $(L_l-L_0)/c=(L_0-L_s)/c=(\Delta L)/c=\tau_0>t_{coh}$, where $c$ and $t_{coh}$ are the speed of the light and the coherence time of the signal or idler photon, respectively. Because of this condition,  the single photon does not interfere with itself and the counting rates of the single detector remain fairly constant. As calculated in Appendix A, the coincidence count rates between two detectors (D1 and D2) as a function of time delay $\tau$ for the modified HOM interference can be expressed as
\begin{equation}
\label{R}
R(\tau)= \frac{1}{4}\int_{0}^{\infty}\int_{0}^{\infty}d\omega_s d\omega_i r(\omega_s,\omega_i,\tau).
\end{equation}
where $r(\omega_s,\omega_i,\tau)$ is the CPD, which reads
\begin{equation}
\label{r}
r(\omega_s,\omega_i,\tau)=|f(\omega_i,\omega_s)(1+e^{-i2\omega_i \tau_0})e^{-i\omega_s (\tau_0+\tau)}-f(\omega_s,\omega_i)(1+e^{-i2\omega_s \tau_0})e^{-i\omega_i (\tau_0+\tau)}|^2.
\end{equation}
where $\tau=2x/c$. $x$ is the position of the beam splitter 3(BS3, 50/50). $f(\omega_s,\omega_i)$ is the two-photon spectral amplitude (TPSA) of the signal and idler photons and its modulo square is the JSI that presents the frequency correlation between the signal and the idler photons. For simplicity, we assume that the TPSA has the exchanging symmetry in our discussions below, i.e., $f(\omega_s,\omega_i)=f(\omega_i,\omega_s)$ . It can be found that if $\tau_0=0$, i.e., $L_l=L_s$, the above equations will deduce to the one of a standard HOM interference \cite{Jin2018}.

Assuming that $\omega_s=\omega_p/2+\Omega_s$ and $\omega_i=\omega_p/2+\Omega_i$, where $\Omega_{s,i}$ is the frequency detuning between the signal (idler) photon and the half of pump center frequency $\omega_p/2$, then Eq.(\ref{r}) can be expressed as the form of the frequency sum $\Omega_{+}=\Omega_s+\Omega_i$ and the frequency difference $\Omega_-=\Omega_s-\Omega_i$,
\begin{eqnarray}
\label{r-pulse}
r(\Omega_+,\Omega_-,\tau)=4|f(\Omega_+,\Omega_-)|^2[1+\cos(\phi+\Omega_+ \tau_0)\cos(\Omega_-\tau_0)-\cos(\phi+\Omega_+ \tau_0)\cos(\Omega_-\tau)\nonumber\\
-\frac{1}{2}\cos(\Omega_-(\tau+\tau_0))-\frac{1}{2}\cos(\Omega_-(\tau-\tau_0))].
\end{eqnarray}
where $\phi=\omega_p\tau_0$. If we define $F(\Omega_{\pm})=|f(\Omega_{\pm})|^2$, its Fourier transform would be
\begin{equation}
G_{\pm}(\tau)=\frac{1}{\sqrt {2\pi}}\int_{-\infty}^{\infty}F(\Omega_{\pm}) e^{i\Omega_{\pm} \tau}d\Omega_{\pm}.
\end{equation}
Integrating Eq.(\ref{r-pulse}) over the entire frequency range, it is found that $R(\tau)$ can be expressed as the functions of $G(\tau)$. We can thus obtain the normalized coincidence count rates,
\begin{equation}
\label{Rn}
R_{N}(\tau)=1+a-bg_-(\tau)-\frac{1}{2}g_-(\tau+\tau_0)-\frac{1}{2}g_-(\tau-\tau_0).
\end{equation}
where $a=\cos(\phi)g_+(\tau_0)g_-(\tau_0)$, $b=\cos(\phi)g_+(\tau_0)$, and $g_{\pm}(\tau)=Re[G_{\pm}(\tau)/G_{\pm}(0)]$. As an example, we take the TPSA as the product of two Gaussian functions, i.e., $f(\Omega_+,\Omega_-)=\exp(-\Omega_+^2/4\sigma_+^2)\exp(-\Omega_-^2/4\sigma_-^2)$, where $\sigma_{\pm}$ denote the linewidth of two functions. In this case, we find that $g_+(\tau_0)g_-(\tau_0)=\exp(-\sigma_+^2\tau_0^2/2)\exp(-\sigma_-^2\tau_0^2/2)$. Because $\tau_0$ must be larger than the two-photon coherence time (determined by $1/\sigma_-$), the second term in Eq.(\ref{Rn}) (independent of $\tau$) tends to be zero whenever the biphotons are frequency correlated ($\sigma_+>\sigma_-$), anti-correlated ($\sigma_+<\sigma_-$) or uncorrelated ($\sigma_+=\sigma_-$). The third term in Eq.(\ref{Rn}) corresponds to the original HOM interference centered at $\tau=0$ and its interference visibility is determined by $b$. If the phase $\phi$ are equal to $\pi/2$ or $3\pi/2$, the visibility will always be zero and not related to the pump linewidth $\sigma_+$. The change of phase $\phi$ can be realized by adjusting the time difference $\tau_0$ between paths $L_l$ and $L_s$ or the pump center frequency $\omega_p$. The final two terms in Eq.(\ref{Rn}) are independent of $\phi$ and correspond to two dips centered at $\tau=\pm \tau_0$ both with an interference visibility of 50\%. The two-side dips are separated by a time interval of $2 \tau_0$.

For a continuous-wave (CW)-pumped and frequency-degenerate SPDC process, we have $\omega_s=\omega_p/2+\Omega$ and $\omega_i=\omega_p/2-\Omega$. We thus get $\Omega_{+}=0$ and $\Omega_-=2\Omega$, then the CPD becomes
\begin{eqnarray}
\label{r-CW}
r_{c}(\tau,\Omega)=4|f(\Omega)|^2|[1+\cos(\phi)\cos(2\Omega\tau_0)-\cos(\phi)\cos(2\Omega\tau)
-\frac{1}{2}\cos(2\Omega(\tau+\tau_0))-\frac{1}{2}\cos(2\Omega(\tau-\tau_0))].
\end{eqnarray}
Integrating Eq.(\ref{r-CW}) over the entire frequency range, the normalized coincidence count rates becomes
\begin{equation}
\label{R-CW}
R_{Nc}(\tau)=1-\cos(\phi)g(2\tau)-\frac{1}{2}g(2(\tau+\tau_0))-\frac{1}{2}g(2(\tau-\tau_0)).
\end{equation}
where $g(\tau)=Re[G(\tau)/G(0)]$. In this case, the interference visibility of the middle interferogram is only determined by the phase $\phi$.

Figure \ref{Fig1}(b) shows the normalized interferogram for the modified HOM interferometer with different values of $\phi$, indicated by the lines of different colors. The middle interferogram is varied for different phases of $\phi$. If the phases $\phi$ are subsequently equal to 0, $\pi/2$(or $3\pi/2$), and $\pi$, we could observe the middle interferogram with the shape of a dip (red), a flat (green) and a peak (purple) around $\tau=0$. The phenomenons where HOM interference pattern around $\tau=0$ changed from a dip to a peak have been observed experimentally in Ref. \cite{PRA1998,PRA2006,OE2017}. The two-side dips always remain no matter how $\phi$ changes. For the pulse-pumped case, the two-side dips will be the same as in the CW-pumped case for the same coherence time of biphotons. The difference is that the visibility of the middle interferogram is now determined by $b$, which can be changed from -1 to 1 by varying both the phase $\phi$ and the pump linewidth $\sigma_+$, as shown in the inset of Fig. \ref{Fig1}(b). Also, the visibility will always be zero and not related to the pump linewidth when the phase $\phi$ is equal to $\pi/2$ or $3\pi/2$. For other fixed phases $\phi$, the visibility decreases as the pump linewidth increases.

Note that the width of all dips or peaks in the modified HOM interferometer is only determined by the two-photon coherence time, which is inversely proportional to the spectral linewidth of biphotons ($\sigma_-$). Additionally, $\tau_0$ must be larger than the coherence time of biphotons, otherwise, the interferogram will tend to be the standard HOM interference as the $\tau_0$ decreases to zero. It can be seen that Eq.(\ref{R-CW}) can be expressed as a function of $g(\tau)$, Fourier transform of $|f(\Omega)|^2$. Therefore, if we write $f(\Omega)=|f(\Omega)|\exp(i \phi(\Omega))$, then the $g(\tau)$ is independent of the phase of $\phi(\Omega)$. It means that these equations are phase-insensitive with regard to the dispersion of the material and are only determined by the spectrum of biphotons. If two materials with identical dispersion are placed in the signal arm and idler arm in a modified or standard HOM interferometer, respectively, all-order dispersion will be canceled, as demonstrated in Ref.\cite{OE2017,OE2021}. Such dispersion cancellation has already been demonstrated experimentally with classical light \cite{Zhou-PR}, however, it can be explained only by the concept of a single-photon wave packet rather than a two-photon one. Therefore, dispersion cancellation in the modified HOM interferometer is one of few phenomena that demonstrate the nonclassicality of the energy-time entangled state for CW-pumped SPDC sources which have strong temporal correlation and frequency anti-correlation ($\phi(\Omega)=\phi(-\Omega)$).

\section{\label{sec:3}Spectrally resolved two-photon interference in a modified HOM interferometer}

%It can be found from Eq.(\ref{Rn}) that the coincidence count rates are determined only by the frequency difference while the spectral information of frequency sum was lost in the time-domain interferogram. It is thus necessary to find the lost information by---

\begin{figure}[th]
\begin{picture}(400,230)
\put(0,0){\makebox(370,220){
\scalebox{0.3}[0.3]{
\includegraphics{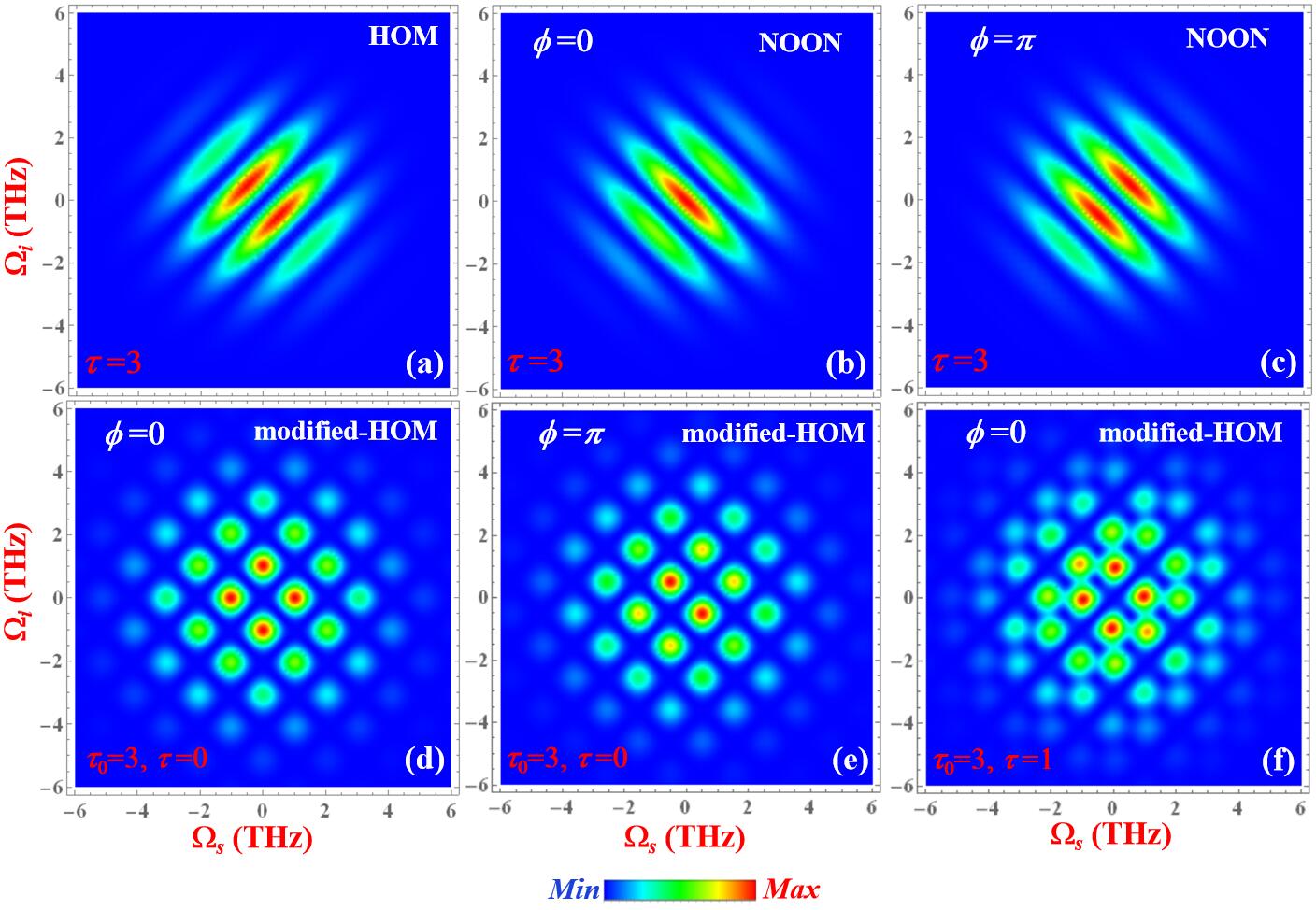}
}}}
\end{picture}
\caption{\label{Fig2}
Spectrally resolved results of the standard HOM interference (a) and N00N state interference at $\phi=0$ (b) and $\phi=\pi$ (c), as a contrast.  The second row represents the spectrally resolved results for the modified HOM interferometer at $\tau_0=3$ with $\phi=0$, $\tau=0$ (d), $\phi=\pi$, $\tau=0$ (e), and $\phi=0$, $\tau=1$ (f). $\sigma_-=\sigma_+=5THz$.}
\end{figure}

Eq.(\ref{Rn}) and Eq.(\ref{R-CW}) are obtained by integrating Eq.(\ref{r-pulse}) and Eq.(\ref{r-CW}) over the entire frequency range. If the output photons are all spectrally resolved so that the frequency integral can be removed, one would obtain the spectrally resolved two-photon interference. This will help us find more spectral information that can not be observed in the time domain \cite{SR-PRA2015,SR-OE2015,SR-PR2020}. Based on the analysis in Part II, the two-side dips from the modified HOM interferometer always remain for a fixed biphoton source no matter how $\phi$ changes, while the middle interferogram becomes flat and its visibility is zero around $\tau=0$ in the case of CW pump with $\phi=\pi/2(3\pi/2)$ or broadband pump. This means that no information about temporal interference can be obtained from the middle interferogram in these two cases. In addition, the middle interferogram in the time domain still keeps flat as the linewidth of the pump pulse continues to increase, as shown in the inset of Fig. \ref{Fig1}(b), indicating that the information of the pulse linewidth will be lost in the temporal interferogram. It is therefore necessary to see whether spectral information can be found through spectrally resolved interference at $\tau=0$. We now come back to review Eq.(\ref{r-pulse}). When $\tau=0$, it can be reduced to
\begin{eqnarray}
\label{r-pulse-1}
r(\Omega_+,\Omega_-,\tau=0)=4|f(\Omega_+,\Omega_-)|^2[1-\cos(\phi+\Omega_+ \tau_0)][1-\cos(\Omega_-\tau_0)].
\end{eqnarray}
It can be seen from Eq.(\ref{r-pulse-1}) that the CPD depends on both the frequency sum and the frequency difference between signal and idler photons, and the JSI is modulated by the time delay $\tau_0$ along both the frequency sum and the frequency difference directions. As an example, we take $\sigma_-=\sigma_+=5THz$, which means that the JSI of the biphotons is frequency un-correlated. In this case, the middle interferogram of the modified HOM interference is flat, and no time-domain interference information can be extracted around $\tau=0$. However, the spectrally resolved interference with a high visibility can be found in Fig. \ref{Fig2}(d)-(f).

Figure \ref{Fig2} shows the spectrally resolved results of the standard HOM interference (a) and N00N state interference (b)-(c), as a contrast (see Appendix C).  Since the standard HOM interference only depends on the frequency difference between signal and idler photons, its spectrally resolved result is phase-independent, and the result at zero delay is a null while the results at other delay $\tau$ can be considered as a cosine modulation on the JSI along the axis of frequency difference (diagonal direction), as shown in Fig.\ref{Fig2}(a). In addition, the spectrally resolved N00N state interference at other delays ($\tau \ne 0$), which is phase-dependent, yielding a cosine modulation on the JSI along the axis of frequency sum (anti-diagonal direction), as shown in Fig.\ref{Fig2}(b)($\phi=0$) and Fig.\ref{Fig2}(c)($\phi=\pi$). The modulation frequency will become much larger as the time delay $\tau$ increases. Both of them have been realized in experiments \cite{SR-PRA2015,SR-OE2015,3-photons2019,3-photons2019,SR-NOON2021}.

The second row in Fig.\ref{Fig2} shows the spectrally resolved results of the modified HOM interference with high visibility. It can be seen that the JSI is indeed modulated by the time delay $\tau_0$ along both frequency sum (anti-diagonal direction) and frequency difference (the diagonal direction) directions and the phase $\phi$ can change the positions of modulations along the directions of frequency sum, as predicted from Eq.(\ref{r-pulse-1}). Note that there is a phase difference of $\pi$ on modulation along the directions of frequency sum between spectrally resolved N00N state interference and the modified HOM interference. This is because the sign of the cosine function contained in the CPD between them is opposite, in spite of the same dependence on frequency sum(see Eq.(\ref{r-pulse-1}) and Eq.(\ref{r+-})). The spectrally resolved results  will become more complicated when both $\tau$ and $\tau_0$ are not equal to zero though the middle interferogram may still be flat, as shown in Fig.\ref{Fig2}(f). Note that the CPD shown in Fig.\ref{Fig2} can be measured quickly in an experiment with a custom very fast single-photon camera, as shown recently in Ref.\cite{Lipka-PRL,Lipka-OL}.

%In this case, it is possible to realize higher dimensional frequency entanglement and more complex engineering of quantum state by changing the time  delay $\tau_0$, $\tau$ and the phase $\phi$.

On the other hand, the middle interferogram in modified HOM interferometer is also flat in the CW-pumped case of $\phi=\pi/2$, or $3\pi/2$. In this case, Eq.(\ref{r-CW}) can be simplified as $r_{c}(\tau=0,\Omega)=4|f(\Omega)|^2|[1-\cos(2\Omega\tau_0)]$. The result is phase-independent and becomes a one-dimensional modulation for the spectrum of the signal or idler photon. To give more clear explanation, it is useful to take the Fourier transform of $r_{c}(\tau,\Omega)$ into $\tilde{r}_c(\tau, T)$, where $T$ is the Fourier conjugate variable to $\Omega$.
\begin{figure}[th]
\begin{picture}(400,200)
\put(0,0){\makebox(370,180){
\scalebox{0.25}[0.25]{
\includegraphics{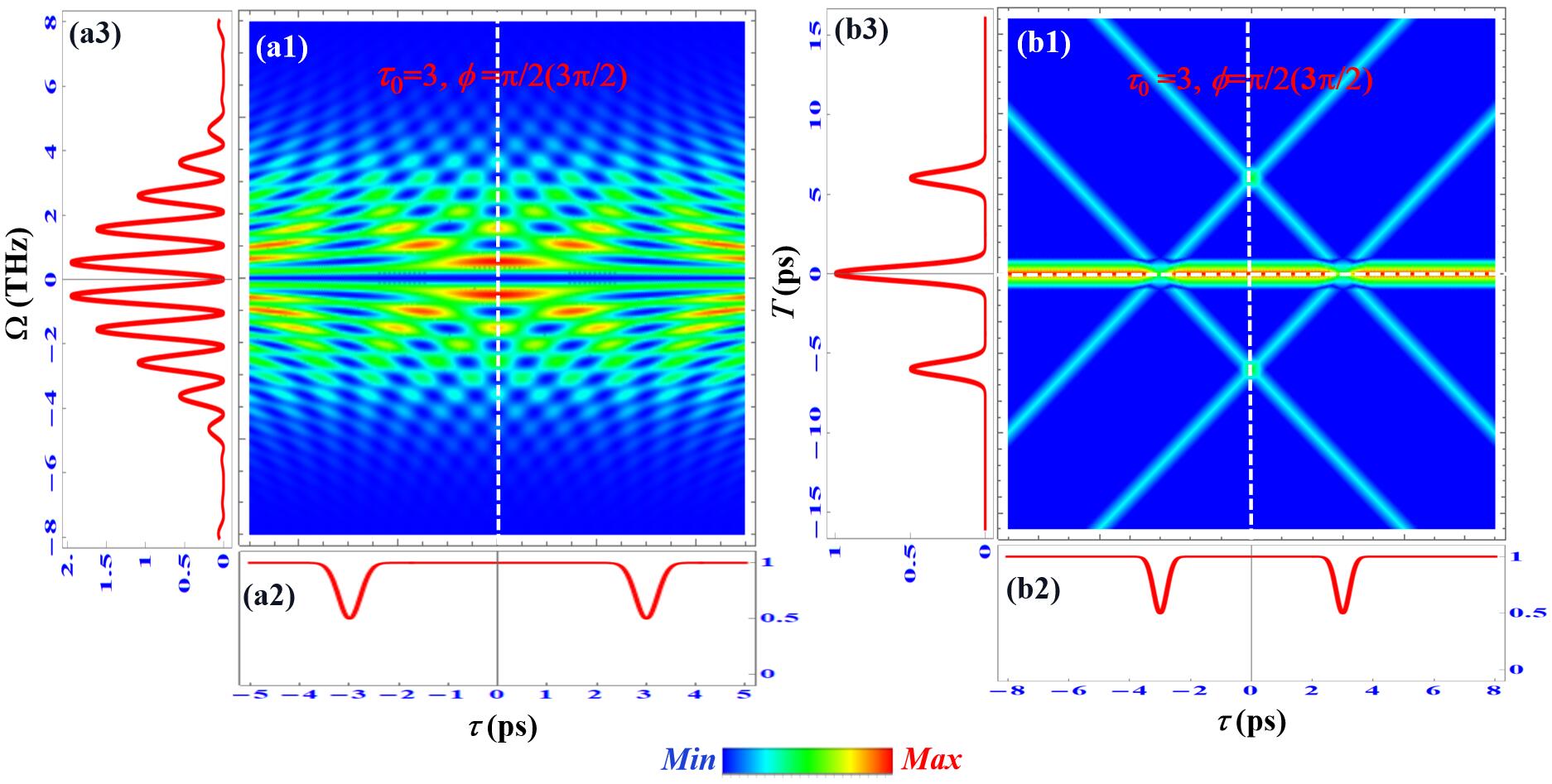}
}}}
\end{picture}
\vspace{0.1cm}
\caption{\label{Fig3}
(a1) The frequency-delay interferogram $r_c(\tau,\Omega)$ in the modified HOM interferometer with $\phi=\pi/2(3\pi/2)$. $\sigma_-=5THz$; $\tau_0=3ps$. Result of integrating the interferogram over $\Omega$, yielding the projection (a2)  of interferogram (a1) in vertical directions. (a3)  Evaluation of $r_c(\tau=0,\Omega)$. (b1) Corresponding time-domain interferogram $|\tilde{r}_c(\tau, T)|$. (b2) Evaluation of $|\tilde{r}_c(\tau, T=0)|$. (b3) Evaluation of $|\tilde{r}_c(\tau=0, T)|$.}
\end{figure}

Figure \ref{Fig3}(a1) shows 2D plot of the frequency-delay interferogram $r_c(\tau,\Omega)$ for the modified HOM interferometer with $\phi=\pi/2(3\pi/2)$. Figure \ref{Fig3}(a2) is the result of integrating the interferogram over $\Omega$, yielding the projection (a2) of interferogram (a1) in the vertical  direction, which is coincident with the temporal interferogram in Fig. \ref{Fig1}(b) at $\phi=\pi/2(3\pi/2)$. There is also no temporal interference information that can be extracted around $\tau=0$. However, the spectrally resolved interference with high visibility can be found in Fig. \ref{Fig3}(a3), which is the result of the evaluation of $r_c(\tau=0,\Omega)$. The result is identical to the projection of Fig. \ref{Fig2}(d) along the axis of the frequency sum, leaving only the frequency difference. Figure \ref{Fig3}(b1) show the corresponding time-domain interferogram $|\tilde{r}_c(\tau, T)|$ . Figure \ref{Fig3}(b2) is the result of the evaluation of $|\tilde{r}_c(\tau, T=0)|$, which is coincident with the temporal interferogram in  Fig. \ref{Fig3}(a2). Figure \ref{Fig3}(b3) is the result of the evaluation of $|\tilde{r}_c(\tau=0, T)|$, which corresponds to Fourier transform of Fig. \ref{Fig3}(a3).

%Also, it should be noted that although there is a bit similarity between the modified HOM interference and the standard HOM interference with a two-interface sample in mathematics, they are two totally different physical processes.

\section{\label{sec:4}Generation and characterization of high-dimensional frequency entanglement}

For a perfectly frequency anti-correlated resource($\sigma_+<<\sigma_-$), the CPD at $\tau=0$ approximates as $|f(\Omega_-)|^2[1-\cos(\Omega_-\tau_0)]$. This is similar to the case of spectrally resolved results in a standard HOM interferometer, where frequency modulation on the JSI along the axis of frequency difference for the frequency anti-correlated entangled resource has been proved to be very useful for generating high-dimensional entanglement in frequency-bin qudits\cite{jin2016,npj2021}. On the other hand, for a perfectly frequency correlated resource($\sigma_+>>\sigma_-$), the CPD at $\tau=0$ approximates as $|f(\Omega_+)|^2[1-\cos(\phi+ \Omega_+\tau_0)]$. In this case, the JSI is phase-dependent and can be modulated by the time delay along the directions of the frequency sum between the signal and idler photons, just like in the case of spectrally resolved N00N state interference\cite{3-photons2019,SR-NOON2021}.

We have noticed that the projection of the JSI along the axis of frequency of the signal or the idler photons in Fig.\ref{Fig2}(d)-(e) will result in the spectra of idler or signal photons with a comb-like structure. It is thus also possible to generate the high-dimensional frequency entanglement with the modified HOM interferometer by changing the time delay $\tau_0$ at $\tau=0$. The numbers of the comb (thus the dimensionality of frequency entanglement) can be adjusted by the time delay $\tau_0$. Also, the phase $\phi$ will change the distributions of the axis of the frequency sum (the anti-diagonal direction) between the signal and idler photons. Similarly, as the time delay $\tau_0$ increases, the modulation frequency becomes much larger, indicating that the higher-dimensional frequency entanglement can be obtained by increasing $\tau_0$.

\begin{figure}[th]
\begin{picture}(400,250)
\put(0,0){\makebox(390,240){
\scalebox{0.32}[0.32]{
\includegraphics{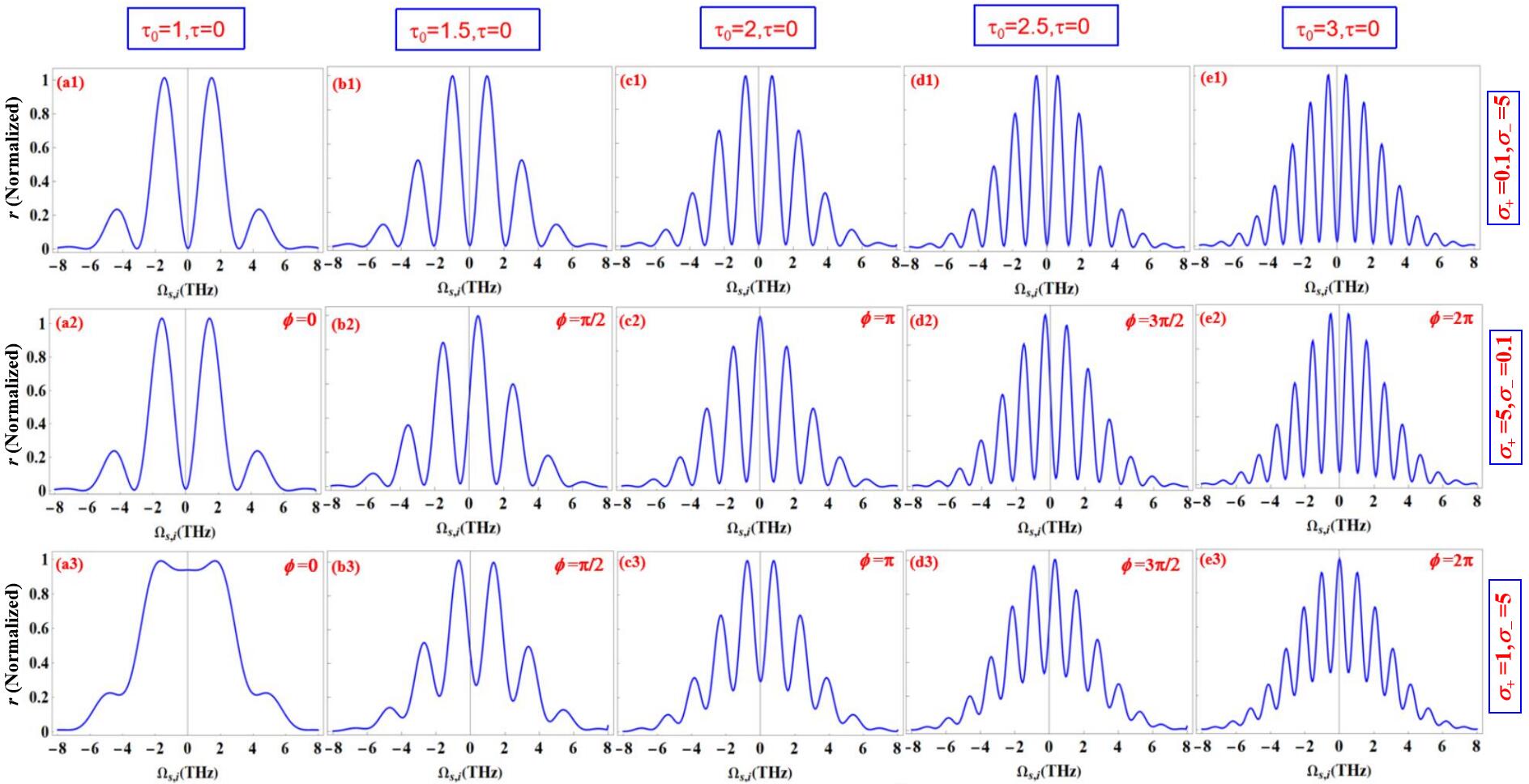}
}}}
\end{picture}
\caption{\label{Fig4}
The simulation of the generation of high-dimensional frequency entanglement with the modified HOM interferometer at $\tau=0$ by changing the time delay $\tau_0$  for different phase $\phi$. The first to the third rows represent the results with a perfectly frequency anti-correlated resource ($\sigma_+=0.1THz, \sigma_-=5THz$), a perfectly frequency correlated resource ($\sigma_+=5THz, \sigma_-=0.1THz$), and a generally frequency anti-correlated resource ($\sigma_+=1THz,\sigma_-=5THz$), respectively. $\tau_0$ is set to be 1ps, 1.5ps, 2ps, 2.5ps, 3ps from the first column(a) to the last column(e), respectively.}
\end{figure}

To certify discussions above, we present some examples in  Fig.\ref{Fig4} on the generation of high-dimensional frequency entanglement with the modified HOM interferometer. The first to the third rows represent the results with a perfectly frequency anti-correlated resource, a perfectly frequency correlated resource, and a generally frequency anti-correlated resource, respectively. It can be seen from the first column (a) to the last column (e) that the numbers of comb increase as the increase of time delay $\tau_0$ . However, the phase of the comb is quite different. In the first row, the frequency of biphotons is perfectly anti-correlated, the JSI is thus phase-independent, and the results of the frequency combs are similar to the case of the standard HOM interferometer \cite{jin2016,npj2021}, where the center of frequency combs is a null.  In the second row, the frequency of biphotons is perfectly correlated, the JSI is thus phase-dependent, and the results of frequency combs are similar to the case of the N00N state interferometer \cite{3-photons2019,SR-NOON2021} but with the opposite phase. In the third row, the frequency of biphotons is generally anti-correlated, and the results of frequency combs are a bit similar to the case of the second row but with the opposite phase and the reduced contrast.

Therefore, it is also possible to generate high-dimensional frequency entanglement with the modified HOM interferometer by modulating the JSI either along the axis of frequency difference for frequency anti-correlated entangled resource or along the axis of the frequency sum for frequency correlated entangled resource by adjusting the time delay $\tau_0$ or $\tau$. In principle, arbitrary-dimensional discrete frequency entanglement can be prepared by tuning the time delay in a modified HOM interferometer. Moreover, the phase of such frequency bins can also be manipulated by adjusting the phase $\phi$. As shown above, the dimensionality of frequency entanglement increase as the increase of time delay $\tau_0$, it is thus possible to characterize directly the generated frequency entanglement with the modified HOM interferometer itself, according to the time interval 2$\tau_0$ between two-side dips appeared in temporal interferogram. It means that the larger the time interval 2$\tau_0$, the higher the dimensionality of frequency entanglement.

%However, in the previous experiment, the JSI can be only modulated along frequency sum in NOON state interference or frequency difference in a standard HOM interference. In is concluded that spectrally resolved results of the modified HOM interferometer combined the advantages of both NOON state interference and standard HOM interference in a single interferometer.

\section{\label{sec:5}CONCLUSIONS}
We have presented a theoretical analysis of the spectrally resolved two-photon interference for the modified HOM interferometer. Our results have shown that the spectrally resolved interference with high visibility in the modified HOM interferometer can be obtained even if its visibility in time domain is zero. Moreover, the coincidence probability density depends on both the frequency sum and the frequency difference between the  signal and idler photons, it is thus phase-dependent and combines both advantages of spectrally resolved HOM interference and N00N state interference. Finally, as a potential application, we have shown that it is possible to generate high-dimensional frequency entanglement by adjusting the time delay and characterize it directly with the two-side dips appeared in temporal interferogram. We hope that these findings could facilitate further  development of spectrally resolved multi-photon interference and its applications in quantum information sciences.

\begin{acknowledgments}
This work has been supported by National Natural Science Foundation of China (12074309, 12074299, 12033007,  61875205, 12103058, 61801458), and the Youth Innovation Team of Shaanxi Universities.
\end{acknowledgments}

\vspace{0.5cm}

\appendix

\section{The coincidence count rates for the modified HOM interferometer derived from  frequency domain}
In this Section, we deduce the equations for the modified HOM interferometer from frequency domain, with the setup shown in Fig.1(a). The idler path length is $L_0$. An unbalanced MZ interferometer is placed in signal arm, and there are two optical paths $L_l$ and $L_s$ such that $L_l-L_0=L_0-L_s=\Delta L$. The two-photon state from a spontaneous parametric down-conversion (SPDC) process can be described as

\begin{equation}
|\Psi\rangle =\int\int d\omega_s d\omega_i f(\omega_s, \omega_i)\hat{a}_s^\dag(\omega_s)\hat{a}_i^\dag(\omega_i)|0\rangle
\end{equation}
where $\omega$ is the angular frequency, and $\hat{a}_{s,i}^\dag$ is the creation operator and the subscripts $s$ and $i$ denote the signal and idler photons from SPDC, respectively. $|0\rangle$ stands for a vacuum state. $f(\omega_s, \omega_i)$ is the two-photon spectral amplitude(TPSA) (also called joint spectral amplitude) of the signal and idler photons.

The detection field operators of detector 1 (D1) and detector 2 (D2) are
\begin{equation}
\hat{E}_1^{(+)}(t_1)= \frac{1}{\sqrt{2\pi}}\int_{0}^{\infty}d\omega_1\hat{a}_1(\omega_1)e^{-i\omega_1t_1}
\end{equation}
\begin{equation}
\hat{E}_2^{(+)}(t_2)= \frac{1}{\sqrt{2\pi}}\int_{0}^{\infty}d\omega_2\hat{a}_2(\omega_2)e^{-i\omega_2t_2}
\end{equation}
where the subscripts 1 and 2 denote the photons detected by D1 and D2, respectively. After MZ interferometer (BS2), the annihilation operator describing the signal photon becomes
\begin{equation}
\hat{a}_{s^{'}}(\omega)= \frac{1}{\sqrt{2}}\hat{a}_s(\omega)(1+e^{-i\omega(L_l-L_s)/c)})
\end{equation}
where $c$ is the speed of the light. The transformation rule of the 50/50 beamsplitter (BS3) is
\begin{equation}
\hat{a}_1(\omega_1)= [\hat{a}_{s^{'}}(\omega_1)+\hat{a}_i(\omega_1)e^{-i\omega_1(L_l-L_0+2x)/c)}]/\sqrt{2}
\end{equation}
\begin{equation}
\hat{a}_2(\omega_2)= [\hat{a}_{s^{'}}(\omega_2)-\hat{a}_i(\omega_2)e^{-i\omega_2(L_l-L_0+2x)/c)}]/\sqrt{2}
\end{equation}
For the sake of simplicity, we set $\tau=2x/c$, $\tau_0=\Delta L/c$ and use the relation $L_l-L_0=L_0-L_s=\Delta L$, then we can rewrite the detection fields as
\begin{equation}
\hat{E}_1^{(+)}(t_1)= \frac{1}{\sqrt{4\pi}}\left[\int_{0}^{\infty}d\omega_1\hat{a}_s(\omega_1)(1+e^{-i2\omega_1 \tau_0}) e^{-i\omega_1 t_1}+\int_{0}^{\infty}d\omega_1\hat{a}_i(\omega_1)e^{-i\omega_1(\tau_0+\tau)}e^{-i\omega_1t_1}\right ]
\end{equation}
\begin{equation}
\hat{E}_2^{(+)}(t_2)= \frac{1}{\sqrt{4\pi}}\left[\int_{0}^{\infty}d\omega_2\hat{a}_s(\omega_2)(1+e^{-i2\omega_2 \tau_0}) e^{-i\omega_2 t_2}-\int_{0}^{\infty}d\omega_2\hat{a}_i(\omega_2)e^{-i\omega_1(\tau_0+\tau)}e^{-i\omega_2t_2}\right ]
\end{equation}
The coincidence count rates between two detectors as a function of delay time $\tau$ can be expressed as
\begin{equation}
R(\tau)= \int \int d t_1 d t_2 \langle \Psi |\hat{E}_1^{(-)}\hat{E}_2^{(-)} \hat{E}_2^{(+)}\hat{E}_1^{(+)}|\Psi \rangle=\int \int d t_1 d t_2 |\langle 0| \hat{E}_2^{(+)}\hat{E}_1^{(+)}|\Psi \rangle|^2
\end{equation}
Consider $ \hat{E}_2^{(+)}\hat{E}_1^{(+)}|\Psi \rangle$, only 2 out of 4 terms exist. The first term is
\begin{eqnarray}
-\frac{1}{4\pi}\int_{0}^{\infty}\int_{0}^{\infty}d\omega_1d\omega_2\hat{a}_s(\omega_1)\hat{a}_i(\omega_2)(1+e^{-i2\omega_1 \tau_0})e^{-i\omega_2 (\tau_0+\tau)} e^{-i\omega_1 t_1}e^{-i\omega_2 t_2} \int_{0}^{\infty}\int_{0}^{\infty} d\omega_s d\omega_i f(\omega_s, \omega_i)\nonumber\\
\times \hat{a}_s^\dag(\omega_s)\hat{a}_i^\dag(\omega_i)|0\rangle
= -\frac{1}{4\pi}\int_{0}^{\infty}\int_{0}^{\infty} d\omega_1 d\omega_2 f(\omega_1, \omega_2)e^{-i\omega_1 t_1}e^{-i\omega_2 t_2}(1+e^{-i2\omega_1 \tau_0})e^{-i\omega_2 (\tau_0+\tau)}|0\rangle
\end{eqnarray}
In this calculation, the relationship of $\hat{a}_s(\omega_1)\hat{a}_s^\dag(\omega_s)=\delta(\omega_1-\omega_s),\hat{a}_i(\omega_2)\hat{a}_i^\dag(\omega_i)=\delta(\omega_2-\omega_i)$ are used.

The second term is
\begin{eqnarray}
\frac{1}{4\pi}\int_{0}^{\infty}\int_{0}^{\infty}d\omega_1d\omega_2\hat{a}_i(\omega_1)\hat{a}_s(\omega_2)(1+e^{-i2\omega_2 \tau_0})e^{-i\omega_1 (\tau_0+\tau)} e^{-i\omega_2 t_2}e^{-i\omega_1 t_1} \int_{0}^{\infty}\int_{0}^{\infty} d\omega_s d\omega_i f(\omega_s, \omega_i)\nonumber\\
\times \hat{a}_s^\dag(\omega_s)\hat{a}_i^\dag(\omega_i)|0\rangle
= \frac{1}{4\pi}\int_{0}^{\infty}\int_{0}^{\infty} d\omega_1 d\omega_2 f(\omega_2, \omega_1)e^{-i\omega_2 t_2}e^{-i\omega_1 t_1}(1+e^{-i2\omega_2 \tau_0})e^{-i\omega_1 (\tau_0+\tau)}|0\rangle
\end{eqnarray}
Combine these two terms:
\begin{eqnarray}
\hat{E}_2^{(+)}\hat{E}_1^{(+)}|\Psi \rangle=&&\frac{1}{4\pi}\int_{0}^{\infty}\int_{0}^{\infty}d\omega_1d\omega_2e^{-i\omega_1 t_1}e^{-i\omega_2 t_2}\nonumber\\
&& \times[f(\omega_2, \omega_1)(1+e^{-i2\omega_2 \tau_0})e^{-i\omega_1 (\tau_0+\tau)}-f(\omega_1, \omega_2)(1+e^{-i2\omega_1 \tau_0})e^{-i\omega_2 (\tau_0+\tau)}]|0\rangle
\end{eqnarray}
Then,
\begin{eqnarray}
\langle \Psi |\hat{E}_1^{(-)}\hat{E}_2^{(-)} \hat{E}_2^{(+)}\hat{E}_1^{(+)}|\Psi \rangle=\left(\frac{1}{4\pi}\right)^2\int_{0}^{\infty}\int_{0}^{\infty}d\omega_1d\omega_2d\omega_1^{'}d\omega_2^{'}e^{-i(\omega_1-\omega_1^{'}) t_1}e^{-i(\omega_2-\omega_2^{'}) t_2}\nonumber\\\times[f^{*}(\omega_2^{'}, \omega_1^{'})(1+e^{-i2\omega_2^{'} \tau_0})e^{-i\omega_1^{'} (\tau_0+\tau)}-f^{*}(\omega_1^{'}, \omega_2^{'})(1+e^{-i2\omega_1^{'} \tau_0})e^{-i\omega_2^{'} (\tau_0+\tau)}]\nonumber\\\times[f(\omega_2, \omega_1)(1+e^{-i2\omega_2 \tau_0})e^{-i\omega_1 (\tau_0+\tau)}-f(\omega_1, \omega_2)(1+e^{-i2\omega_1 \tau_0})e^{-i\omega_2 (\tau_0+\tau)}]
\end{eqnarray}
Finally,
\begin{eqnarray}
R(\tau)= \int \int d t_1 d t_2 \langle \Psi |\hat{E}_1^{(-)}\hat{E}_2^{(-)} \hat{E}_2^{(+)}\hat{E}_1^{(+)}|\Psi \rangle=\frac{1}{4}\int_{0}^{\infty}\int_{0}^{\infty}d\omega_1d\omega_2d\omega_1^{'}d\omega_2^{'}\delta(\omega_1-\omega_1^{'})\delta(\omega_2-\omega_2^{'})\nonumber\\\times
[f^{*}(\omega_2^{'}, \omega_1^{'})(1+e^{-i2\omega_2^{'} \tau_0})e^{-i\omega_1^{'} (\tau_0+\tau)}-f^{*}(\omega_1^{'}, \omega_2^{'})(1+e^{-i2\omega_1^{'} \tau_0})e^{-i\omega_2^{'} (\tau_0+\tau)}]
\nonumber\\\times[f(\omega_2, \omega_1)(1+e^{-i2\omega_2 \tau_0})e^{-i\omega_1 (\tau_0+\tau)}-f(\omega_1, \omega_2)(1+e^{-i2\omega_1 \tau_0})e^{-i\omega_2 (\tau_0+\tau)}]
\end{eqnarray}
In above calculation, the relationship of $\delta(\omega-\omega_{'})=\frac{1}{2\pi}\int_{-\infty}^{\infty}e^{i(\omega-\omega^{'})t}dt$ is used. $f^*$ is the complex conjugate of $f$. If the $f$ is symmetric, i.e., $f(\omega_1, \omega_2)=f(\omega_2, \omega_1)$ , we can further rewrite the above equation as
\begin{eqnarray}
\label{R12}
R(\tau)= \frac{1}{4}\int_{0}^{\infty}\int_{0}^{\infty}d\omega_1d\omega_2|f(\omega_1, \omega_2)|^2|(1+e^{-i2\omega_2 \tau_0})e^{-i\omega_1 (\tau_0+\tau)}-(1+e^{-i2\omega_1 \tau_0})e^{-i\omega_2 (\tau_0+\tau)}|^2
\end{eqnarray}
In order to introduce less variables, Eq.(\ref{R12}) can be rewritten as
\begin{equation}
R(\tau)= \frac{1}{4}\int_{0}^{\infty}\int_{0}^{\infty}d\omega_s d\omega_i r(\omega_s,\omega_i,\tau).
\end{equation}
This is Eq.(\ref{R}) in the main context.

\section{The coincidence count rates for the modified HOM interferometer derived from the time domain}
In order to have a comprehensive understanding on the modified HOM interferometer, we now need to Fourier transform Eq.(\ref{r}) back into the time representation, yielding to the JTI
\begin{eqnarray}
\label{FT-r}
|\Psi(t_s,t_i)|^2=&&|A_1(t_s+(\tau+\tau_0), t_i)-A_2(t_s, t_i+(\tau+\tau_0))+A_3(t_s+(\tau+\tau_0), t_i+2\tau_0)\nonumber\\
&&-A_4(t_s+2\tau_0, t_i+(\tau+\tau_0))|^2.
\end{eqnarray}
where $\Psi$ is called as effective two-photon wave function. $A$ denotes the quantum-mechanical probability amplitude for the coincidence detection event and can be obtained from the Fourier transform of the TPSA at the coordinates marked in Eq.(\ref{FT-r}). It can be seen from Eq.(\ref{FT-r}) that there are four  temporal biphoton amplitudes, centered at ($\tau+\tau_0,0$),  ($0,\tau+\tau_0$),  ($\tau+\tau_0,2\tau_0$),  ($2\tau_0,\tau+\tau_0$), respectively, and these amplitudes can be conveniently represented by Feynman-type diagrams as shown in \cite{PRA1998}. After taking module square of Eq.(\ref{FT-r}), the JTI has 16 terms,
\begin{eqnarray}
\label{FT-16}
|\Psi(t_s,t_i)|^2=&&|A_1|^2+|A_2|^2+|A_3|^2+|A_4|^2+A_1^*A_3+A_1A_3^*+A_2^*A_4+A_2A_4^*\nonumber\\
&&-A_1^*A_4-A_1A_4^*-A_2^*A_3-A_2A_3^*-A_1^*A_2-A_1A_2^*-A_3^*A_4-A_3A_4^*.
\end{eqnarray}
The coincidence count rates between two detectors as a function of time  delay $\tau$ can be rewritten as
\begin{equation}
R(\tau)=\int_{-\infty}^{\infty}\int_{-\infty}^{\infty}dt_s dt_i |\Psi(t_s, t_i)|^2
\end{equation}
As an example, we take the TPSA as the product of two Gaussian functions, that is
\begin{equation}
\label{f}
f(\omega_s, \omega_i)=\exp\left(- \frac{(\omega_s+\omega_i-\omega_p)^2}{4\sigma_{+}^2}\right)\exp\left(- \frac{(\omega_s-\omega_i)^2}{4\sigma_{-}^2}\right)
\end{equation}
Taking the Fourier transform of Eq.(\ref {f}) at the coordinates marked in Eq.(\ref {FT-r}) and integrating the first four terms of Eq.(\ref {FT-16}) over all the time range, we can obtain a normalized factor $8\pi\sigma_{+}\sigma_{-}$. Similarly, integrating from the fifth to the eighth, from the ninth to the twelfth,  and from  the last four terms of Eq.(\ref {FT-16}), we can obtain the normalized results, which correspond to the second, the third and the last two terms of Eq.(\ref {Rn}), respectively. We therefore obtain the same result from the time domain as Eq.(\ref {Rn}) derived from the frequency domain.

\section{The coincidence count rates for a standard HOM and N00N state interferometer}
In order to compare the results of a standard HOM and N00N state interferometer with the modified HOM interferometer, we give the coincidence count rates for a standard HOM and N00N state interferometer as follows, (see details in \cite{Jin2018})

\begin{equation}
\label{R+-}
R_{\pm}(\tau)\sim \int_{0}^{\infty}\int_{0}^{\infty}d\omega_s d\omega_i r_{\pm}(\omega_s,\omega_i,\tau).
\end{equation}
where 
\begin{equation}
\label{r+}
r_{+}(\omega_s,\omega_i,\tau)=|f(\omega_i,\omega_s)(e^{-i\omega_i \tau}+1)(e^{-i\omega_s \tau}+1)+f(\omega_s,\omega_i)(e^{-i\omega_s \tau}-1)(e^{-i\omega_i \tau}-1)|^2,
\end{equation}
\begin{equation}
\label{r-}
r_{-}(\omega_s,\omega_i,\tau)=|f(\omega_i,\omega_s)e^{-i\omega_s \tau}-f(\omega_s,\omega_i)e^{-i\omega_i \tau}|^2.
\end{equation}
$``-"$ and $``+"$ corespond to a standard HOM and N00N state interferometer, respectively. If the $f$ is symmetric, i.e., $f(\omega_s, \omega_i)=f(\omega_i, \omega_s)$ , we can further rewrite the above equation as
\begin{equation}
\label{r+-}
r_{\pm}(\omega_s,\omega_i,\tau)=|f(\omega_s,\omega_i)|^2[1\pm \cos(\omega_s \pm \omega_i)\tau].
\end{equation}

\end{document}